\documentclass[fleqn,twoside]{article}
\usepackage{espcrc2}

\usepackage{graphicx}

\title{AN ALTERNATIVE INTERPRETATION OF X(3872)}

\author{Kamal K. Seth\address[MCSD]{Department of Physics and Astronomy, 
        Northwestern University\\Evanston, IL, 60208, USA}}
       
\begin{document}

\begin{abstract}
It is proposed that X(3872) is a vector glueball mixed with neighboring vector states of charmonium.
\end{abstract}

\maketitle

The Belle Collaboration [1] recently reported the observation of a narrow state, dubbed X(3872) in the decay of a large sample of 152 million $B$ mesons, $B^\pm\to K^\pm\mathrm{X}$, with X decaying into $\pi^+\pi^- J/\psi$.  The observation was confirmed by the CDF II Collaboration [2] in $p\bar{p}$ collisions at 1.8 TeV, and subsequently by the D$\O$ Collaboration [3] and the BaBar Collaboration [4].  Table 1 lists the reported masses and widths.

\begin{table*}[htb]
\caption{Measured Masses and Widths of X(3872)}
\begin{tabular}{l|c|c|l}
\hline \hline
Collaboration & Mass & Width & Reaction\\
 & (MeV) & (MeV), 90\% CL & \\
\hline
Belle [1] & $3872.0\pm0.6\pm0.5$ & $\le2.3$ & $B\to K\mathrm{X},\mathrm{X}\to\pi^+\pi^-J/\psi$\\ 
CDF II [2] & $3871.3\pm0.7\pm0.3$ & $\le4.9\pm0.7$ & $p\bar{p}\to\mathrm{X}+\mathrm{any};\mathrm{X}\to\pi^+\pi^-J/\psi$\\ 
D$\O$ [3] & $3871.8\pm3.1\pm3.0$ & $\le17\pm3$ & $p\bar{p}\to\mathrm{X}+\mathrm{any};\mathrm{X}\to\pi^+\pi^-J/\psi$\\ 
BaBar [4] & $3873.4\pm1.4$ & $\le3.1\pm0.2$ & $B\to K\mathrm{X},\mathrm{X}\to\pi^+\pi^-J/\psi$\\ 
\hline
Average & $3871.92\pm0.31$ & $\le2.3$ \\
\hline \hline
\end{tabular}
\end{table*}

The error weighted average mass is
\begin{equation}
M(\mathrm{X}(3872)) = 3871.9\pm0.5 \; \mathrm{MeV}
\end{equation}
The widths are all consistent with the experimental resolution of the respective measurements, so that
\begin{equation}
\Gamma(\mathrm{X}(3872))<2.3\;\mathrm{MeV},\qquad 90\%\mathrm{CL}
\end{equation}
In all four measurements $\psi'(2S)$ decaying into $\pi^+\pi^- J/\psi$ is observed rather strongly, with the corresponding X(3872) being a small fraction.  In terms of the product branching ratios,
\begin{equation}
\frac{\mathcal{B}(B\to K \mathrm{X})\times \mathcal{B}(\mathrm{X} \to \pi^+\pi^- J/\psi)}{\mathcal{B}(B\to K \psi')\times \mathcal{B}(\psi' \to \pi^+\pi^- J/\psi)}
\end{equation}
$$ \begin{array}{l}=(6.3\pm1.4)\%\;(\mathrm{Belle})\\=(8.6\pm1.9)\%\;(\mathrm{BaBar})\end{array}$$
Several attempts have been made to search for X(3872) decays in final states other than $\pi^+\pi^- J/\psi$ [1,5,6,7,8].  While all such searches are handicapped by poor statistics, the overall conclusion is that no positive signals have been observed in any, except perhaps in the decay X$\to\omega J/\psi$ [7].  The 90\% confidence limits established are listed in Table 2 as ratios of the observed decay to $\pi^+\pi^- J/\psi$, together with the corresponding ratios for $\psi'$ decays [9].

The unique characteristics of X(3872) have given rise to a number of theoretical suggestions for its nature.  These essentially fall into two classes: (a) X(3872) is a charmonium state, and (b) because its mass is very close to $(M(D^0)+M(\bar{D}^{0*})=2M(D^0)+(M(D^{*0})-M(D^0))=3871.3\pm1.0$ MeV [9], it is a weakly bound molecule.  Eichten, Lane, and Quigg [10], and Barnes and Godfrey [11], have examined the charmonium options in detail.  The charmonium states in play are $1^{3,1}D_J$ and $2^{3,1}P_J$.  In nearly all potential model and lattice calculations the $2^{3,1}P_J$ states are predicted to have masses 50--100 MeV larger than 3872 MeV, and the masses of $1^{3,1}D_J$ states are predicted to be lower than 3872 MeV, but by smaller amounts.  Both Barnes and Godfrey [11] and Eichten et al [10] have calculated total widths of the $1^{3,1}D_J$ and  $2^{3,1}P_J$ states.  BG find that only $^3D_3$, $^{3,1}D_2$, and $^{3,1}P_1$ states have small enough widths to be compatible with the $<2.3$ MeV measured width of X(3872).  ELQ find  $^3D_3$, $^{3,1}D_2$, and $^3P_2$ to be narrow, but $^3P_1$ (and perhaps $^1P_1$ also) to be too wide.  Distinguishing between these possibilities is not going to be easy.  In the meanwhile, for the $D^0\bar{D}^{0*}$ molecule interpretation of X(3872), Swanson proposes $J^{PC}=1^{++}$ [12] and Turnqvist proposes $J^{PC}=1^{++}$ and $0^{-+}$ [13].  Of course, we should note that with the currently available mass measurements
\begin{equation}
M(\mathrm{X(3872)})-(M(D^0)+M(D^{0*})) = 0.6\pm1.1\;\mathrm{MeV}
\end{equation}
the possibility exists that X(3872) may not be the bound state of $D^0$ and $\bar{D}^{0*}$.  Hopefully, better mass measurements will soon provide the definitive verdict on whether the molecular interpretation is at all feasible.

In this letter, we propose an alternative to the current explanations of X(3872).  We propose that X(3872) is mainly a vector glueball which has acquired a small admixture of vector $|c\bar{c}>$ states in its neighborhood.  It is this admixed $|c\bar{c}>(1^{--})$ component which is responsible for the observed decay of X(3872) to $\pi^+\pi^- J/\psi$, and for the fact that both CDF [2] and D$\O$ [3] note that X(3872) decay in this channel resembles that of $\psi'(3686)$ decay in all detailed characteristics.  Further, we show that none of the existing measurements in which X(3872) is not observed are in contradiction with this proposition.

\begin{table}[bht]
\caption{Ratios $R(Res.\to f.s.)\equiv \mathcal{B}(Res.\to f.s.)$ $/\mathcal{B}(Res.\to\pi^+\pi^- J/\psi)$, where $Res.= $X(3872) or $\psi'(3686)$.}
\begin{tabular}{l|c|c}
\hline
\multicolumn{1}{c|}{Final State ($f.s.$)} & $R(\mathrm{X}\to f.s.)$ & $R(\psi'\to f.s.)$ \\
\multicolumn{1}{c|}{[Ref]} & 90\% C.L. & PDG04 \cite{2004partlist}\vspace{1pt}  \\
\hline
\vspace{1pt} [1] $\gamma \chi_{c1}$ & $<0.89$ & 0.29 \\
\vspace{1pt} [7] $\gamma \chi_{c2}$ & $<1.1$ & 0.29 \\
\vspace{1pt} [6] $\eta J/\psi$ & $<0.6$ & 0.10 \\
\vspace{1pt} [7] $\pi^0\pi^0 J/\psi$ & $<0.74$ & 0.5 \\
\vspace{1pt} [7] $\gamma J/\psi$ & $<0.40$ & -- \\
\vspace{1pt} [7] $\omega J/\psi$ & $0.8\pm0.3$ & -- \\
\vspace{1pt} [5] $D^0\bar{D}^0$ & $<4.4$ & -- \\
\vspace{1pt} [5] $D^+D^-$ & $<2.9$ & -- \\
\hline
\end{tabular}
\end{table}

The possibility of X(3872) as a vector glueball is based on the well-known lattice calculation of glueball masses by Morningstar and Peardon [14].  They predict a $J^{PC}=1^{--}$ 3-gluon vector glueball at a mass $M=3850\pm50\pm190$ MeV.  The pure glueball would not couple to a photon, and have no $(e^+e^-)$ width.  However, it would certainly mix with neighboring charmonium vectors, $\psi'(3686)$, $\psi''(3770)$ and $\psi'''(4040)$, which are generally considered to be $2^3S_1$, $1^3D_1$ and $3^3S_1$ $|c\bar{c}>$ states, respectively.  In prior discussions of $0^{++}$ glueballs, it is admitted that the mixing matrix element between a $0^{++}$ glueball state and $0^{++}$ ($n\bar{n}$, $s\bar{s}$) states is not known, and it has been introduced as an adjustable parameter [15].  Reasonably efficient mixing of a $|gg>_{0^{++}}$ state with $|n\bar{n},s\bar{s}>_{0^{++}}$ states leading to $30-40\%$ amplitude admixtures between $|gg>$ and $|n\bar{n},s\bar{s}>$ states separated by up to 200 MeV, has been advocated.  In the charmonium sector, no such estimates of mixing between glueballs and $|c\bar{c}>$ states have been made.  However, in their recent paper, Eichten, Lane and Quigg [10] have considered in detail the consequences of mixing of charmonium states near open flavour threshold.  The mixing induced through open $c\bar{c}$ channels is found to be large.  For example, they find that $\psi''(3770)$ wave function is:
\begin{eqnarray*}
|\psi''(3770)> = 0.69|1^3D_1>+0.17e^{0.8i\pi}|2^3S_1> \\ +0.10e^{0.9i\pi}|2^3D_1>+\,\mathrm{smaller...}
\end{eqnarray*}
A similar admixture of $\psi'(2^3S_1)$ in the three gluon vector glueball with $M=3872$ MeV would lead to $\psi'(\mathrm{amplitude})^2\sim3\%$.  We show that such an admixture could explain all the existing measurements of the decays of X(3872).

It is difficult to estimate $\mathcal{B}(\mathrm{X}\to\pi^+\pi^-J/\psi)/\mathcal{B}(\psi'\to\pi^+\pi^-J/\psi)$ from the measured ratio of the product branching ratios in eq. 3, which average to $7\pm1\%$, because the production of both X and $\psi'$ involves weak decay of $B$ mesons.  Similarly, production of X and $\psi'$ in $p\bar{p}$ annhilation at TeV energies presumably involves many indirect routes.  However, both these problems are avoided in $e^+e^-$ production of X and $\psi'$.  In a recent search for the production of X(3872) in ISR mediated $e^+e^-$ annihilation, CLEO [16] has established 90\% confidence upper limit
\begin{equation} 
\Gamma_{ee} \mathcal{B}(\mathrm{X}\to\pi^+\pi^-J/\psi) <8.3\;\mathrm{eV}
\end{equation} 
which leads to the 90\% confidence upper limit
\begin{equation} 
\frac{\Gamma_{ee}(\mathrm{X}) \mathcal{B}(\mathrm{X}\to\pi^+\pi^-J/\psi)}{\Gamma_{ee}(\psi') \mathcal{B}(\psi'\to\pi^+\pi^-J/\psi)} < \frac{8.3\;\mathrm{eV}}{672\;\mathrm{eV}},
\end{equation}
$$\mathrm{or}\;<0.0124$$
using the measured width and branching ratio for $\psi'$ [9]. We now examine the implications of this result.

Let us consider two extreme scenarios.  Consider an amplitude $f$ for the $\psi'(2^3S_1)$ content in X(3872).  Its formation in ISR mediated $e^+e^-$ annihilation will be $\Gamma_{ee}(\mathrm{X})=f^2\Gamma_{ee}(\psi')$.  If the glueball component in the wave function of X leads to negligably small contribution to its total width, the branching ratio $\mathcal{B}(\mathrm{X}\to\pi^+\pi^- J/\psi)=\mathcal{B}(\psi'\to\pi^+\pi^- J/\psi)$.  From eq. 6 we therefore obtain
\begin{equation}
f^2<0.0124,\quad\mathrm{or}\quad f<11\%
\end{equation}
The other extreme is that the glueball component of the wave function of X(3872) leads to other (non $\pi^+\pi^-J/\psi$) decays, and its total width increases from that of $\psi'$ ($\Gamma(\psi')$=0.3 MeV) to the present experimental limit of $\Gamma$(X)$<$2.3 MeV, or by a factor 7.7.  Then $\mathcal{B}(\mathrm{X}\to\pi^+\pi^- J/\psi)=\mathcal{B}(\psi'\to\pi^+\pi^- J/\psi)/7.7$.  From eq. 6 we then obtain
\begin{equation}
f^2<(7.7)\times0.0124,\quad\mathrm{or}\quad f<30\%
\end{equation}
Neither of the limits in eq. 7 and 8 appear to be improbable\footnote{We note that CLEO [16] obtained the limit $f<20\%$, on the basis of entirely different assumptions about $B$ decays.}, considering that we have not only $\psi'(3686)$ in the neighborhood of X(3872) but at least two other vector states of charmonium $\psi''(3770)$ and $\psi'''(4040)$ not far away [9].  However, only an actual calculation can determine what level of vector $|c\bar{c}>$ admixture can be expected in a vector glueball state at 3872 MeV.  As mentioned earlier, the mixing matrix elements are not known, and without their knowledge reliable predictions are not possible.

As noted in Table 2, it is known that the ratios 
\begin{equation}R\left(\frac{\mathcal{B}(\psi'\to\gamma\chi_{1,2})}{\mathcal{B}(\psi'\to\pi^+\pi^-J/\psi)}\right) = 0.29,\end{equation}
\begin{equation}R\left(\frac{\mathcal{B}(\psi'\to\eta J/\psi)}{\mathcal{B}(\psi'\to\pi^+\pi^-J/\psi)}\right) = 0.10,\end{equation}
\begin{equation}R\left(\frac{\mathcal{B}(\psi'\to\pi^0\pi^0 J/\psi)}{\mathcal{B}(\psi'\to\pi^+\pi^-J/\psi)}\right) = 0.5,\end{equation}

 so that the present limits for the corresponding ratios for the decays of X(3872) listed in \linebreak Table 2 are automatically satisfied by our model of X(3872).  In our model, the decays X$\to\gamma J/\psi$ and X$\to\omega J/\psi$ would not be allowed.  We can not, of course, obtain the limits for X$\to D^0\bar{D}^0,D^+D^-$ from $\psi'$.

D$\O$ has made a detailed comparison of several characteristics of the events they observe for X(3872)$\to\pi^+\pi^-J/\psi$ and $\psi'(3686)\to\pi^+\pi^-J/\psi$.  They find  near perfect agreement between the two for the distributions of $p_T$, rapidity, $\theta_\pi$, $\theta_\mu$, decay length, and isolation [3].  CDF has made a study of the lifetime and finds that the long lived fractions of the $\pi^+\pi^-J/\psi$ decays are also identical for X(3872) and $\psi'(3686)$ [2].  These measurements provide additional support for our contention that the X(3872)$\to\pi^+\pi^-J/\psi$ events owe their origin to the vector $|c\bar{c}>$ content of X(3872).

The obvious shortcoming of our model for X(3872) is our lack of knowledge of the decay channels and the partial widths for the vector glueball component of the X(3872) wave function.  The closest analogy we can make is to the three gluon annihilation widths of $J/\psi$ ($\approx$60 keV) and $\psi'$ ($\approx100$ keV).  If the decay of a vector glueball with three gluons can be modeled after these decays of $J/\psi$ and $\psi'$, the contribution of the glueball component of X(3872) to its total width is likely to be at a sub-MeV level.  Further, it would appear that the decay of the three gluon state to the four quark $D\bar{D}$ system, $|ggg>(1^{--})\to |c\bar{n}>+|\bar{c}n>$, is likely to be also very weak.  In any case, it has not been observed so far.

It has been noted that the proposal that X(3872) is a $D^0\bar{D}^{0*}$ molecule can be eliminated by the observation of its decay into $\pi^0\pi^0 J/\psi$. Similarly, our proposal that X(3872) is a vector can be disproved by the unambiguous observation of its decay into $\gamma J/\psi$ or $\omega J/\psi$.

Our proposal for the glueball nature of X(3872) raises several theoretical questions, and we hope that lattice calculations and lattice-motivated models for glueballs will address them, and test the viability of our model.

This research was supported by the U.S. Department of Energy.

\end{document}